\documentclass[twocolumn,showpacs,nopreprintnumbers,amsmath,pra]{revtex4}
\usepackage{amsmath,amssymb}
\usepackage{graphicx}

\begin{document}

\title{Coherent population trapping in quantized light
field}

\author{A. V. Ta\u\i chenachev, A. M. Tuma\u\i kin, and V. I. Yudin}
\affiliation{Novosibirsk State University, Pirogova 2, Novosibirsk 630090, Russia\\
Institute of Laser Physics SB RAS, Lavrentyeva 13/3, Novosibirsk 630090, Russia}
%\author{A. M. Tuma\u\i kin}
%\affiliation{Novosibirsk State University, Pirogova 2, Novosibirsk 630090, Russia}
%\affiliation{Institute of Laser Physics SB RAS, Lavrentyeva 13/3, Novosibirsk 630090, Russia}
%\author{V. I. Yudin}
%\affiliation{Novosibirsk State University, Pirogova 2, Novosibirsk 630090, Russia}
%\affiliation{Institute of Laser Physics SB RAS, Lavrentyeva 13/3, Novosibirsk 630090, Russia}

\date{\today}

\begin{abstract}
A full quantum treatment of coherent population trapping (CPT) is
given for a system of resonantly coupled atoms and
electromagnetic field. We develop a regular analytical method of
the construction of generalized dark states (GDS). It turns out
that GDS do exist for all optical transitions $F_g\to F_e$,
including bright transitions $F\to F+1$ and $F''\to F''$ with
$F''$ a half-integer, for which the CPT effect is absent in a
classical field. We propose an idea to use an optically thick
medium with a transition $F''\to F''$ with $F'' \ge 3/2$ a
half-integer as a ''quantum filter'', which transmits only a
quantum light.
\end{abstract}


\maketitle

%{\bf 1. Intro.}

Coherent population trapping (CPT) for atoms interacting with a resonant field
is now well-known \cite{Alz,Ar1} and widely used in various branches of atomic
and laser physics. Let us list several examples: nonlinear optics of resonant
media \cite{NLO}, nonlinear high-resolution spectroscopy \cite{DSres},
high-precision metrology (atomic clocks and magnetometers) \cite{DSmetr}, laser
cooling \cite{VSCPT}, atom optics and interferometry \cite{AT}, and quantum
information processing \cite{DSPol,lightstorage}. Originally the CPT theory was
developed for a three-state model \cite{Ar2,Gr}. A generalization to multilevel
atomic systems was done by Hioe and Carrol \cite{Hioe} without regard for the
relaxation. In particular, in \cite{Hioe} sufficient conditions for the
existence of dark (non-coupled with light) states at the two-photon resonance
have been derived. For the resonant excitation by a classical elliptically
polarized field from the atomic ground state with angular-momentum degeneracy,
necessary and sufficient conditions for the occurrence of CPT and the explicit
form of dark states have been established in our works \cite{sm,CPT_inv}. With
allowance made for the radiative relaxation, all optical transitions $F_g\to
F_e$ ($F_g,F_e$ are the total angular momenta of the ground $(g)$ and excited
$(e)$ states) were classified into two classes depending on the occurrence of
CPT: (i) dark transitions $F\to F-1$ with $F$ arbitrary and $F'\to F'$  with
$F'$ an integer; (ii) bright transitions $F\to F+1$ with $F$ arbitrary and
$F''\to F''$ with $F''$ a half-integer. In the later case CPT is absent.

It was shown in \cite{dalton82} that the laser-field fluctuations of certain
type destruct the dark state, and the CPT effect disappears. This rule,
however, is not general. For instance, we have found in \cite{taich94} that the
dark state can arise upon the interaction with an arbitrarily fluctuating
monochromatic radiation field. In particular, it has been shown that the CPT
effect occurs even if the photon spin coherence is completely absent.

Recent outstanding theoretical works on the problem of dark-state polariton
\cite{mazets96,DSPol}  and the experimental demonstration of the optical
information storage in resonant atomic media  \cite{lightstorage} attracted an
attention to the problem on the most general form of dark state in the
``atoms+field'' system. It should be noted that the experimental results
obtained in \cite{lightstorage} are of great importance in view of their
possible applications to the storage of essentially quantum information encoded
in nonclassical states of light. In this connection, the following statement of
the problem looks quite reasonable: what happens with the CPT existence
conditions and with the form of dark states, when the resonant field is
described by the creation and annihilation operators instead of c-number
amplitudes? From this point of view the dark-state polariton should be a
particular case of the general solution, when one field mode is classical (i.e.
it is in a coherent state), while another mode is quantal. Moreover, as it was
argued in \cite{cornellnature}, it is quite possible that, for the long-term
storage of quantum information in an atomic medium, it will be necessary to use
ultracold atomic ensembles (e.g. Bose-Einstein condensates). In this case,
atoms can conveniently be described by the secondary quantized amplitudes, i.e.
by the creation and annihilation operators of Bose or Fermi types
\cite{juzeliunas02}.

In the present paper coherent population trapping in an atomic ensemble with
optical transition $F_g\to F_e$ upon the resonance  interaction with a
quantized monochromatic running wave is considered. We develop a regular method
of the construction of generalized dark states (GDS) for arbitrary angular
momenta $F_g,F_e$. It is shown that for the dark transitions ($F\to F-1$ with
$F$ arbitrary and $F'\to F'$  with $F'$ an integer) the CPT effect occurs at
arbitrary numbers of atoms and photons, representing a direct generalization of
the results obtained for the classical elliptically polarized field \cite{sm}.
Unexpectedly, it turns out that in the quantized field CPT can appear even for
the bright transitions ($F\to F+1$ with $F$ arbitrary and $F''\to F''$  with
$F''$ a half-integer), for which the dark states are absent in the classical
field \cite{sm}. In this case, however, we have certain limitations on the
number of photons in one of two polarization modes for the transitions $F''\to
F''$  with $F''$ a half-integer, and on the numbers of both atoms and photons
for the transitions $F\to F+1$ with $F$ arbitrary.

%{\bf 2. Problem statement}

%We consider the resonant interaction of an atomic ensemble with a quantized
%monochromatic running wave. The field drives an optical transition $F_g\to F_e$, where $F_g$ and $F_e$
%are the total angular momenta in the ground and excited states, respectively.
Hereinafter, we assume that atoms obey to either Bose-Einstein or Fermi-Dirac
statistics. The total Hamiltonian of the ``atoms+field'' system  is a sum of
three summands:
$\widehat{H}=\widehat{H}_{a}+\widehat{H}_{ph}+\widehat{H}_{a-ph}$. The
Hamiltonian of free atoms can be written as
\begin{equation}\label{A}
\widehat{H}_{a}=\hbar\omega_0\sum_{\mu_e}\hat{c}^{\dag}_{\mu_e}\hat{c}_{\mu_e},
\end{equation}
where $\omega_0$ is the optical transition frequency,
$\hat{c}^{\dag}_{\mu_e}$ and $\hat{c}_{\mu_e}$ are the operators
of creation and annihilation of an atom in the excited-state
Zeeman substate $|F_e,\mu_e\rangle$ ($-F_e\le \mu_e\le F_e$). If
the quantization axis is directed along the wavevector, then the
Hamiltonian of free transversal photons reads
\begin{equation}\label{Ph}
\widehat{H}_{ph}=\hbar\omega\sum_{s=\pm 1}\hat{\alpha}^{\dag}_{s}\hat{\alpha}_{s},
\end{equation}
where $\omega$ is the field frequency,
$\hat{\alpha}_{s}$($\hat{\alpha}^{\dag}_{s}$) is the bosonic
annihilation (creation) operator of a photon with the right
($s=+1$) or left ($s=-1$) circular polarization.
% its temporal dependence is given by $\hat{\alpha}_{s}=\hat{\alpha}_{s}(0)\exp(-i\omega t)$.
In the rotating wave approximation the
dipole coupling between atoms and field is described by the operator
$\widehat{H}_{a-ph}=\widehat{V}+\widehat{V}^{\dag}$, where
\begin{equation}\label{V}
\widehat{V}=\hbar\Omega\sum_{\mu_e\mu_gs}\hat{c}^{\dag}_
{\mu_e}C^{F_e\mu_e}_{F_g\mu_g,1s}\hat{b}_{\mu_g}\hat{\alpha}_{s}\,.
\end{equation}
Here $\Omega$ is the single-photon Rabi frequency, $C^{F_e\mu_e}_{F_g\mu_g,1s}$
are the Clebsch-Gordan coefficients, and
$\hat{b}^{\dag}_{\mu_e}$($\hat{b}_{\mu_e}$) is the operator of creation
(annihilation) of an atom in the ground-state Zeeman substate
$|F_g,\mu_g\rangle$ ($-F_g\le \mu_g \le F_g$).

Our goal is to find the generalized dark states $|NC\rangle$ of the system
``atoms+field'', which nullify the interaction operator:
\begin{equation}\label{Dark}
\widehat{V}\,|NC\rangle=0\,.
\end{equation}
In addition the state $|NC\rangle$ should be stable with respect to the radiative relaxation, i.e.
GDS should not contain atoms in the excited state. Formally this condition can be expressed as
\begin{equation}\label{Dark2}
\widehat{H}_{a}\,|NC\rangle=0\,,
\end{equation}
in line with eq.(\ref{A}), where just the excited-state operators
$\hat{c}_{\mu_e}$ are present. Combining  eq.(\ref{Dark}) and eq.(\ref{Dark2}),
one can see that $\widehat{H}_{a-ph}\,|NC\rangle=0$.

%{\bf 3. Analysis of possible variants.}

Taking into account the angular-momentum selection rules, we see that the
generic scheme of the light-induced transitions is split into two independent
chains for arbitrary optical transition $F_g\to F_e$. There exist three types
of such chains \cite{sm}. For the first $\Lambda$-type of chains the number of
coupled ground-state substates is greater than those in the excited state by
one. The $\Lambda$-chains are realized (see in Fig.1{\em a,b}) in the
transitions $F\to F-1$ with $F$ arbitrary and $F'\to F'$  with $F'$ an integer.
For the second $V$-type of chains the number of coupled ground-state substates
is less than those in the excited state by one. The $V$-chains are realized
(see in Fig.1{\em b,c}) for the transitions  $F\to F+1$ with $F$ arbitrary and
$F'\to F'$  with $F'$ an integer. For the third type of chains the number of
coupled ground-state substates is equal to those in the excited state. These
chains will be referred to as $N$-chains. The $N$-chains are realized (see in
Fig.1{\em d}) for the transitions $F''\to F''$ with $F''$ a half-integer. Let
us develop a procedure of the GDS construction for each type of chains
separately.

{\bf\small The $\Lambda$-chains.} Consider the $\Lambda$-chain consisting of
$L$ links. For the sake of convenience we re-enumerate the substates as it is
shown in Fig.1{\em a} (lower panel). The ground-state substates have odd
numbers, while those in the excited state have even numbers. The coupling
operator (\ref{V}) projection on this $\Lambda$-chain sub-space can be written
as
\begin{equation}\label{V_lam}
\widehat{V}_{\Lambda}=\hbar\Omega\sum_{j=1}^{L}\hat{c}^{\dag}_{2j}(G^{2j}_{2j-1}\hat{b}_{2j-1}\hat{\alpha}_{+1}+
G^{2j}_{2j+1}\hat{b}_{2j+1}\hat{\alpha}_{-1}),
\end{equation}
where $G^k_l$ denote the corresponding Clebsch-Gordan coefficients (compare
with eq.(\ref{V})).

We will need in the following construction:
\begin{equation}\label{Psi}
\widehat{\Psi}_{NC}=\sum_{j=0}^{L}\widehat{A}_{2j+1}\hat{b}^{\dag}_{2j+1}\,,
\end{equation}
which is a superposition of the atomic ground-state creation
operators with operator coefficients $\widehat{A}_{2j+1}$. These
coefficients are functions of the photon annihilation operators
$\hat{\alpha}_{\pm 1}$. Using the standard commutation (bosons) or
anticommutation (fermions) rules, one can obtain:
\begin{eqnarray}\label{V_Psi}
&&\widehat{V}_{\Lambda}\widehat{\Psi}_{NC}=\pm\widehat{\Psi}_{NC}\widehat{V}_{\Lambda}+\\
&&\qquad\hbar\Omega\sum_{j=1}^L\hat{c}^{\dag}_{2j}(G^{2j}_{2j-1}\hat{\alpha}_{+1}\widehat{A}_{2j-1}+
G^{2j}_{2j+1}\hat{\alpha}_{-1}\widehat{A}_{2j+1})\,,\nonumber
\end{eqnarray}
where the sign $+/-$ corresponds to bosons/fermions. The coefficients
$\widehat{A}_{l}$ are chosen in such a way that the second term in the r.h.s.
of eq.(\ref{V_Psi}) becomes zero. In the other words  $\widehat{A}_{l}$ obey
the recurrent operator equations:
\begin{equation}\label{rek}
G^{2j}_{2j-1}\hat{\alpha}_{+1}\widehat{A}_{2j-1}+
G^{2j}_{2j+1}\hat{\alpha}_{-1}\widehat{A}_{2j+1}=0\,.
\end{equation}
These equations always have a solution, because the number of equations in
(\ref{rek}) equals $L$, while the number of the coefficients  $\widehat{A}_{k}$
equals $L+1$. The solution can be written in the form of positive powers of the
annihilation operators $\hat{\alpha}_{\pm 1}$:
\begin{eqnarray}\label{prod}
&&\widehat{A}_{2j+1}=\\
&&(-1)^j(\hat{\alpha}_{+1})^j(\hat{\alpha}_{-1})^{L-j}\left[\prod_{q=1}^{j}G^{2q}_{2q-1}\right]
\left[\prod_{q'=j+1}^LG^{2q'}_{2q'+1}\right] \, ,\nonumber
\end{eqnarray}
where $j=0,...,L$, and the convention $\prod_{k+1}^{k}\equiv 1$ is used. With
this choice of the coefficients $\widehat{A}_{2j+1}$ it follows from
eq.(\ref{V_Psi}) that
\begin{equation}\label{fund}
\widehat{V}_{\Lambda}\widehat{\Psi}_{NC}=\pm\widehat{\Psi}_{NC}\widehat{V}_{\Lambda}\,.
\end{equation}
This fundamental relationship allows us to construct the dark state
$|NC\rangle_{\Lambda}$ for the $\Lambda$-chain in the form:
\begin{equation}\label{NC2}
|NC\rangle_{\Lambda}=(\widehat{\Psi}_{NC})^n\widehat{\Phi}\{\hat{\alpha}^{\dag}\} |0\rangle\,,
\end{equation}
where $n$ is arbitrary non-negative number, the functional
$\widehat{\Phi}\{\hat{\alpha}^{\dag}\}$ depends exclusively on the photon
operators $\hat{\alpha}^{\dag}_{\pm 1}$, and $|0\rangle$ is the vacuum state of
atoms and photons. Indeed, using eq.(\ref{fund}), we get
\begin{equation}\label{dok}
\widehat{V}_{\Lambda}|NC\rangle_{\Lambda} =(\pm 1)^n(\widehat{\Psi}_{NC})^n\widehat{V}_{\Lambda}
\widehat{\Phi}\{\hat{\alpha}^{\dag}\}|0\rangle =0\,.
\end{equation}
The later equality follows from the fact that the coupling operator
$\widehat{V}_{\Lambda}$ contains the annihilation operators $\hat{b}_{k}$,
which commute with $\widehat{\Phi}\{\hat{\alpha}^{\dag}\}$ and give zero,
acting on the vacuum state. It is also obvious that the state
$|NC\rangle_{\Lambda}$ obeys the equation (\ref{Dark2}), because the
excited-state operators $\hat{c}^{\dag}_{k}$ do not enter in the construction
(\ref{Psi}).

The exponent $n$ in eq.(\ref{NC2}) means the number of atoms participating in
the formation of the dark state. In the Bose case this number can be arbitrary,
while in the Fermi case we have just two possibilities $n=0,1$, since
$(\widehat{\Psi}_{NC})^2=0$. However, this fact does not lead to the principal
limitation on the number of atoms in an ensemble. Indeed, taking into account
the translational degrees of freedom, for example in the momentum
representation, we can write the more general from of GDS:
\begin{equation}\label{NC3}
|NC\rangle_{\Lambda}=\left[\prod_{\bf p}(\widehat{\Psi}_{NC}({\bf p}))^{n_{\bf
p}}\right]\widehat{\Phi}\{\hat{\alpha}^{\dag}\}|0\rangle\,,
\end{equation}
where the creation operators $\hat{b}^{\dag}_{k}({\bf p})$ and the basic
construction $\widehat{\Psi}_{NC}({\bf p})$ are labeled by the momentum ${\bf
p}$. The total number of atoms is $\sum_{\bf p}n_{\bf p}$, and it can be
arbitrary in the Fermi case as well. It should be noted that an arbitrary
superposition of the dark states (\ref{NC3}) will be dark, according to our
definitions (\ref{Dark}) and (\ref{Dark2}).

It is interesting that the basic construction  $\widehat{\Psi}_{NC}$ can be
obtained from the results of our early work \cite{sm} in a formal way. To do
this one has to use the explicit form of the dark states in the classical
field, substituting formally the Zeeman substate wavefunctions  by the atom
creation operators $\hat{b}^{\dag}_{m}$ and the circular field components by
the photon annihilation operators $\hat{\alpha}_{\pm 1}$. In this sense the
obtained GDS $|NC\rangle_{\Lambda}$ can be viewed as a direct generalization of
the results of the paper \cite{sm} to the quantized field and many-particle
atomic system.

As an example, we demonstrate here how does the $m$-fold excited dark-state
polariton \cite{DSPol} emerge in our approach. We consider the simplest
$\Lambda$-chain (i.e. $L=1$) with the operator construction (\ref{Psi}) given
by $\widehat{\Psi}_{NC}=G_3^2\hat{\alpha}_{-1}\hat{b}^{\dag}_{1}-
G_1^2\hat{\alpha}_{+1}\hat{b}^{\dag}_{3}$. The functional
$\widehat{\Phi}\{\hat{\alpha}^{\dag}\}$ in eq.(\ref{NC2}) is taken in the form:
\begin{equation}\label{polar}
\widehat{\Phi}\{\hat{\alpha}^{\dag}\}= {\cal N}_0\,(\hat{\alpha}_{-1}^{\dag})^m
\exp\{Z\hat{\alpha}_{+1}^{\dag}\}\,,
\end{equation}
where ${\cal N}_0$ is a normalization constant. When acting on the vacuum state
$|0\rangle$, the operator (\ref{polar}) generates a coherent state with
amplitude $Z$ in the mode with circular polarization $(+1)$ and $m$ photons in
the mode with circular polarization $(-1)$.  The direct calculations by formula
(\ref{NC2}) with $G_1^2=G_3^2$ yield the result that coincides (except for
notation) with the $m$-fold excited state $|D,m\rangle$ of the dark-state
polariton \cite{DSPol}.

{\bf\small The $N$-chains.} Consider the Zeeman substates, for which the
light-induced transitions form the $N_{\pm}$-chains, as it is shown in
Fig.1{\em d} (two lower panels) for the transitions $F''\to F''$ with $F''$ a
half-integer. Then the coupling operator projections on these subspaces are
written as
\begin{eqnarray}\label{V_N+}
&&\widehat{V}_{N_{+}}=\widehat{V}_{\Lambda}+\hbar\Omega G^{2L+2}_{2L+1}\,\hat{c}^{\dag}_{2L+2}
\hat{\alpha}_{+1}\hat{b}_{2L+1}\,,\\
\label{V_N-}&&\widehat{V}_{N_{-}}=\widehat{V}_{\Lambda}+\hbar\Omega G^{0}_{1}\,\hat{c}^{\dag}_{0}
\hat{\alpha}_{-1}\hat{b}_{1}\,,
\end{eqnarray}
where $\widehat{V}_{\Lambda}$ denotes the part of the interaction operator
connecting the substates into the maximal $\Lambda$-chain with $L$ links. This
$\Lambda$-chain is marked in Fig.1{\em d} by double lines. The explicit form of
$\widehat{V}_{\Lambda}$ corresponds to eq.(\ref{V_lam}). The second terms in
the operators (\ref{V_N+}) and (\ref{V_N-}) describe the interaction via the
outermost excited-state Zeeman substate, which is connected with just one
ground-state substate. This coupling is shown in Fig.1{\em d} by single line.

For the main contribution $\widehat{V}_{\Lambda}$, in line with eq.(\ref{Psi})
and  eq.(\ref{prod}), we construct the operator $\widehat{\Psi}_{NC}$, which
obeys the equation (\ref{fund}). Then, for the $N_{+}$-chain the dark state
$|NC\rangle_{N_{+}}$, obeying the equation
$\widehat{V}_{N_{+}}|NC\rangle_{N_{+}}=0$, has the form:
\begin{equation}\label{NC_N+}
|NC\rangle_{N_{+}}=(\widehat{\Psi}_{NC})^n(\hat{\alpha}^{\dag}_{+1})^m \widehat{\Phi}\{\hat{\alpha}^{\dag}_{-
1}\}|0\rangle,\; (m\le L),
\end{equation}
where the generic operator functional
$\widehat{\Phi}\{\hat{\alpha}^{\dag}_{-1}\}$ depends only on the
left-polarized photon creation operators
$\hat{\alpha}^{\dag}_{-1}$. Thus, the number of the left-polarized
photon can be arbitrary. It is obvious that (\ref{NC_N+})
nullifies the operator $\widehat{V}_{\Lambda}$ (see
eq.(\ref{dok})). The limitation on the number of the
right-polarized photons  $m$ in eq.(\ref{NC_N+}) is connected with
the necessity to nullify the additional contribution $\hbar\Omega
G^{2L+2}_{2L+1}\,\hat{c}^{\dag}_{2L+2}
\hat{\alpha}_{+1}\hat{b}_{2L+1}$ in r.h.s. of eq.(\ref{V_N+}). The
condition $m\le L$ follows from the fact that the operator
construction $\widehat{\Psi}_{NC}$ (\ref{Psi}) contains the
creation operator $\hat{b}^{\dag}_{2L+1}$ with the coefficient
$\widehat{A}_{2L+1}\propto\hat{\alpha}^L_{+1}$ (see
eq.(\ref{prod})).

In much the same way, we find for theÿ $N_{-}$-chain:
\begin{equation}\label{NC_N-}
|NC\rangle_{N_{-}}=(\widehat{\Psi}_{NC})^n(\hat{\alpha}^{\dag}_{-1})^m \widehat{\Phi}\{\hat{\alpha}^{\dag}_{+
1}\}|0\rangle,\; (m\le L),
\end{equation}
where the number of the left-polarized photons is limited by $m\le L$.

Strictly speaking, in eqs.(\ref{NC_N+}),(\ref{NC_N-}) we have to put the number
of atoms $n=0,1$, since the $N$-chains correspond to half-integer angular
momenta, i.e. to fermions. Because of this, analogously to (\ref{NC3}), we
present the dark state in more general form:
\begin{equation}\label{NC_NP}
|NC\rangle_{N_{\pm}}=\left[\prod_{\bf p}(\widehat{\Psi}_{NC}({\bf p}))^{n_{\bf
p}}\right](\hat{\alpha}^{\dag}_{\pm 1})^m
\widehat{\Phi}\{\hat{\alpha}^{\dag}_{\mp 1}\} |0\rangle.
\end{equation}
Now the number of atoms $\sum_{\bf p}n_{\bf p}$ can be arbitrary.

Evidently, for the $N$-chains the $m$-fold excited dark-state polariton
$|D,m\rangle$ \cite{DSPol} can be constructed. However, contrary to the
$\Lambda$-chains, here the number of the quantized mode excitations is limited
by the length of chain $(m\le L)$.

It is interesting to note that if the number of quanta in the
weak mode $m$ exceeds the length of $N$-chain $(m > L)$, then the
dark state is absent and the process of the spontaneous
scattering of photons takes place. We could expect that in the
course of the light pulse propagation in the medium this process
will proceed until the number of quanta in the weak  mode
decreases to the required level $(m\le L)$. After that the pulse
will propagate without scattering, because GDS will be formed.
Thus, the medium of atoms with the transition $F''\to F''$ with
$F'' \ge 3/2$ a half-integer can serve as a ''quantum filter''.
In this case inside the medium an entangled state of atomic
ensemble and finite number of quanta of the weak mode is formed.
At the entrance from the optically thick medium, blocking the
strong mode by a polarization device, we will have the pulse with
a finite number of quanta, i.e. the quantum light.

{\bf\small The $V$-chains.} Consider the Zeeman substates, for which the
light-induced transitions form the $V$-chain, as it is shown in Fig.1{\em c}
(lower panel). Then the coupling operator projection on this subspace
% $\widehat{V}_{V}$
can be written as
\begin{equation}\label{V_V}
\widehat{V}_{V}=\widehat{V}_{\Lambda}+\hbar\Omega (
G^{0}_{1}\,\hat{c}^{\dag}_{0}
\hat{\alpha}_{-1}\hat{b}_{1}+G^{2L+2}_{2L+1}\,\hat{c}^{\dag}_{2L+2}
\hat{\alpha}_{+1}\hat{b}_{2L+1}),
\end{equation}
where $\widehat{V}_{\Lambda}$ denotes the part of the interaction operator
connecting the substates into the maximal $\Lambda$-chain with $L$ links. This
$\Lambda$-chain is marked in Fig.1{\em c} by double lines. The explicit form of
$\widehat{V}_{\Lambda}$ corresponds to eq.(\ref{V_lam}). The two additional
terms in the operator (\ref{V_V}) describe the interaction via the outermost
excited-state Zeeman substates, which are connected with just one ground-state
substate. This coupling is shown in Fig.1{\em c} by single line.

Again, for the main contribution $\widehat{V}_{\Lambda}$, in line with
eq.(\ref{Psi}) and  eq.(\ref{prod}), we construct the operator
$\widehat{\Psi}_{NC}$, which obeys the equation (\ref{fund}). Then, for the
$V$-chain the dark state can be written in the form:
\begin{equation}\label{NC_V}
|NC\rangle_{V}=\widehat{\Psi}_{NC}\,(\hat{\alpha}^{\dag}_{+1})^m (\hat{\alpha}^{\dag}_{-1})^{m'}|0\rangle,\;
(m,m'\le L).
\end{equation}
Here, contrary to the $N$-chains, the numbers of photons of both polarizations
are limited $(m,m'\le L)$. This is because we have to nullify the two
additional terms in r.h.s. of eq.(\ref{V_V}). Furthermore, the state
(\ref{NC_V}) is non-trivial (i.e. it differs from zero and contains photons) if
and only if $(m+m')>L$.

As an example, we consider the $V$-chain with one $\Lambda$-link, i.e. $L=1$.
In this case $\widehat{\Psi}_{NC}=G_3^2\hat{\alpha}_{-1}\hat{b}^{\dag}_{1}-
G_1^2\hat{\alpha}_{+1}\hat{b}^{\dag}_{3}$, and $m=m'=1$. Then using
eq.(\ref{NC_V}), we obtain the expression:
$$|NC\rangle_{V}=(G_3^2\hat{\alpha}^{\dag}_{+1}\hat{b}^{\dag}_{1}-
G_1^2\hat{\alpha}^{\dag}_{-1}\hat{b}^{\dag}_{3})|0\rangle\,,$$ where the atom
and field variables are entangled. Here the quantum entanglement is precisely
the reason leading to the non-trivial dark states on the bright optical
transitions.

It should be noted that the dark states (\ref{NC_V}) are realized for just one
atom. It can be seen from the calculation:
$$(\widehat{\Psi}_{NC})^n(\hat{\alpha}^{\dag}_{+1})^m
(\hat{\alpha}^{\dag}_{-1})^{m'}|0\rangle=0,\quad (n\ge 3;\;m,m'\le L).$$ This
equality follows from the simple count the numbers of creation and annihilation
operators in each term. In the case of two atoms the state
$(\widehat{\Psi}_{NC})^2(\hat{\alpha}^{\dag}_{+1})^m
(\hat{\alpha}^{\dag}_{-1})^{m'}|0\rangle$ is trivial or (at $m=m'=L$) it does
not contain photons.

%{\bf 4. Conclusion.}

Concluding, we have developed the analytical method, which allows
us to construct a wide class of GDS for the chains of
$\Lambda,V,N$-types. These chains of the dipole connected Zeeman
substates appear under the consideration of the resonance
interaction of the quantized plane-wave modes with optical
transitions $F_g \to F_e$. The generalized dark states exist for
arbitrary field frequency $\omega$, they immune to the radiative
relaxation, and their mathematical structure corresponds, in
general case, to quantum states entangled in atomic and field
variables (see eqs.(\ref{Psi}),(\ref{prod})). The obtained results
lead to the following classification of the optical transitions.
(i) The transitions $F\to F-1$ with arbitrary $F$ and $F'\to F'$
with $F'$ an integer, where the $\Lambda$-chains are realized.
Here GDS exist for arbitrary quantum state of the field and for
arbitrary number of atoms. (ii) The transitions $F''\to F''$ with
$F''$ a half-integer, where the $N$-chains are realized. In this
case GDS occur at arbitrary number of atoms, but the number of
photons of the right (or left) circular polarization is limited.
(iii) Transitions $F\to F+1$, where the $V$-chains are realized.
Here we found just the ultra-quantum GDS for one atom interacting
with the limited numbers of photons of both right and left
polarizations.

We would like to stress that in the cases (ii) and (iii) the ground-state dark
states are absent in the classical light field. It is possible that the
obtained GDS do not exhaust all dark states in the ``atoms+field'' system. In
particular, the problem on the existence of a many-particle dark state for the
transitions  $F\to F+1$ is still open. Note also that all the explicit
expressions for GDS can be easily re-formulated for  distinguishable particles.
To do this one has to use the Zeeman wavefunctions instead of the creation
operators $\hat{b}^{\dag}_k$ in the basic construction (\ref{Psi}).

We have found the generalized dark states for the given frequency
$\omega$ and wave-vector ${\bf k}$, then these states can be
marked by $\omega$ and ${\bf k}$: $|NC(\omega,{\bf k})\rangle$.
It is obvious that an arbitrary superposition and even an
incoherent mixture of the dark states corresponding to different
$\omega$ and ${\bf k}$ will obey the conditions (\ref{Dark}) and
(\ref{Dark2}), i.e. they will be dark states.

The obtained results are  of principal significance for the CPT theory, quantum
optics, and quantum informatics in resonant atomic media.

This work is supported by a grant INTAS-01-0855 and by RFBR ( grants
05-02-17086 and 04-02-16488).

% \newpage

\begin{figure*}[h]\centerline{\scalebox{0.5}{\includegraphics{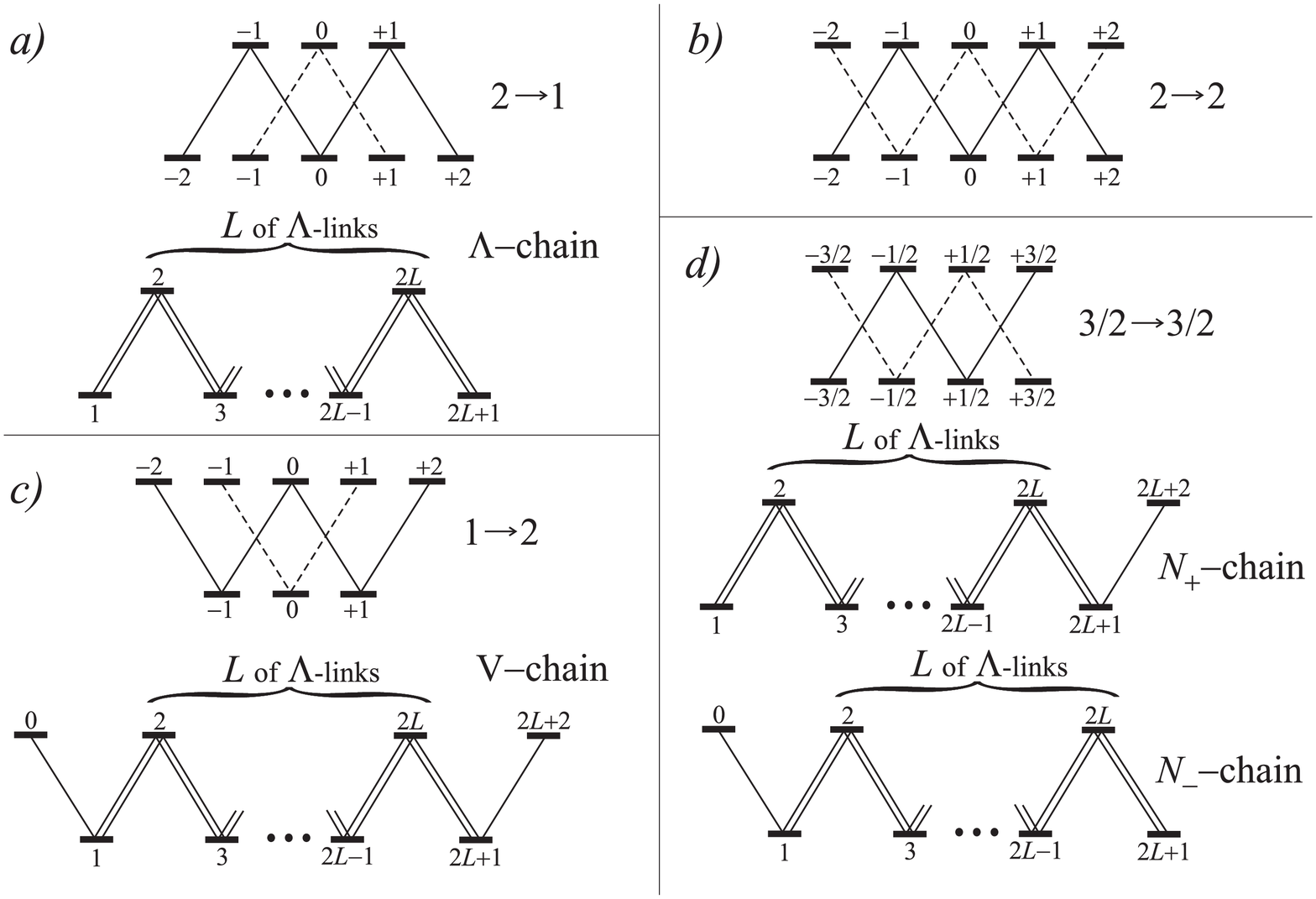}}}
\caption{The light-induced transitions between Zeeman substates in a resonant
plane wave : {\em a)} transitions $F\to F-1$ with $2\to 1$ as an example (upper
panel), the $\Lambda$-type chain (lower panel); {\em b)} transitions $F'\to F'$
($F'$ is an integer) with $2\to 2$ as an example; {\em c)} transitions $F\to
F+1$ with $1\to 2$ as an example (upper panel), the $V$-type chain (lower
panel); {\em d)} transitions $F''\to F''$ ($F''$ is a half-integer) with
$3/2\to 3/2$ as an example (upper panel), the $N_{\pm}$-chains (two lower
panels).} \label{fig1}
\end{figure*}

\end{document}